\documentclass[conference]{defs/IEEEtran}
\IEEEoverridecommandlockouts

\usepackage{cite}
\usepackage{amsmath,amssymb,amsfonts}
\usepackage{mathtools}
\usepackage{amsthm}
\usepackage{algorithm}
\usepackage{algpseudocode} 
\usepackage{etoolbox}
\usepackage{graphicx,color}
\usepackage{textcomp}
\usepackage[dvipsnames,rgb]{xcolor}
\usepackage{float}
\usepackage{tikz}
\usetikzlibrary{arrows.meta, positioning, shapes.geometric,decorations.pathreplacing}
\usetikzlibrary{patterns}
\usepackage{siunitx}
\usepackage{tabularx}
\usepackage{colortbl}
\usepackage{multirow}
\usepackage{array}
\usepackage{acronym}

\usepackage{subcaption}
\usepackage{booktabs}
\usepackage{hyperref} 
\usepackage[capitalize,noabbrev]{cleveref}
\usepackage[textsize=tiny]{todonotes}
\usepackage{latexsym}
\usepackage{gensymb}
\usepackage{enumitem}
\usepackage{esvect}
\usepackage{comment}
\usepackage[normalem]{ulem}
\usepackage{soul}

\input{defs/macros}
\theoremstyle{plain}

\theoremstyle{definition}

\theoremstyle{remark}

\definecolor{nyuviolet}{RGB}{87, 6, 140}

\def\BibTeX{{\rm B\kern-.05em{\sc i\kern-.025em b}\kern-.08em
    T\kern-.1667em\lower.7ex\hbox{E}\kern-.125emX}}

\usepackage[acronyms,nonumberlist,nopostdot,nomain,nogroupskip,toc=false]{glossaries}
\glsdisablehyper

\newacronym{quic}{QUIC}{Quick UDP Internet Connections}
\newacronym{3gpp}{3GPP}{3rd Generation Partnership Project}
\newacronym{adc}{ADC}{Analog to Digital Converter}
\newacronym{dac}{DAC}{Digital to Analog Converter}
\newacronym{if}{IF}{Intermediate Frequency}
\newacronym{sma}{SMA}{SubMiniature Version A}
\newacronym{cot}{CoT}{Commercial Off-the-Shelf}
\newacronym{aod}{AoD}{Angle of Departure}
\newacronym{lo}{LO}{local oscillator }

\newacronym{5g}{5G}{5th generation}

\newacronym{aimd}{AIMD}{Additive Increase Multiplicative Decrease}
\newacronym{am}{AM}{Acknowledged Mode}
\newacronym{amc}{AMC}{Adaptive Modulation and Coding}
\newacronym{aqm}{AQM}{Active Queue Management}
\newacronym{awgn}{AGWN}{Additive White Gaussian Noise}
\newacronym{afd}{AFD}{Austin Fire Department}
\newacronym{balia}{BALIA}{Balanced Link Adaptation}
\newacronym{bdp}{BDP}{Bandwidth-Delay Product}
\newacronym{bf}{BF}{Beamforming}
\newacronym{cc}{CC}{Congestion Control}
\newacronym{cdf}{CDF}{Cumulative Distribution Function}
\newacronym{cn}{CN}{Core Network}
\newacronym{cqi}{CQI}{Channel Quality Information}
\newacronym{cp}{CP}{Control Plane}
\newacronym{csirs}{CSI-RS}{Channel State Information - Reference Signal}
\newacronym{dc}{DC}{Dual Connectivity}
\newacronym{dce}{DCE}{Direct Code Execution}
\newacronym{dci}{DCI}{Downlink Control Information}
\newacronym{dl}{DL}{Downlink}
\newacronym{dmr}{DMR}{Deadline Miss Ratio}
\newacronym{dmrs}{DMRS}{DeModulation Reference Signal}
\newacronym{e2e}{E2E}{End-to-End}
\newacronym{ecn}{ECN}{Explicit Congestion Notification}
\newacronym{edf}{EDF}{Earliest Deadline First}
\newacronym{enb}{eNB}{evolved Node Base}
\newacronym{epc}{EPC}{Evolved Packet Core}
\newacronym{es}{ES}{Edge Server}
\newacronym{fdma}{FDMA}{Frequency Division Multiple Access}
\newacronym{fdd}{FDD}{Frequency Division Duplexing}
\newacronym[firstplural=Radio Access Technologies (RATs)]{rat}{RAT}{Radio Access Technology}
\newacronym{fs}{FS}{Fast Switching}
\newacronym{ftp}{FTP}{File Transfer Protocol}
\newacronym{gnb}{gNB}{Next Generation Node Base}
\newacronym{harq}{HARQ}{Hybrid Automatic Repeat reQuest}
\newacronym{hetnet}{HetNet}{Heterogeneous Network}
\newacronym{hh}{HH}{Hard Handover}
\newacronym{hol}{HOL}{Head-of-Line}
\newacronym{ia}{IA}{Initial Access}
\newacronym{imt}{IMT}{International Mobile Telecommunication}
\newacronym{iot}{IoT}{Internet of Things}
\newacronym{los}{LOS}{Line of Sight}
\newacronym{lte}{LTE}{Long Term Evolution}
\newacronym{m2m}{M2M}{Machine to Machine}
\newacronym{mac}{MAC}{Medium Access Control}
\newacronym{mc}{MC}{Multi-Connectivity}
\newacronym{mcs}{MCS}{Modulation and Coding Scheme}
\newacronym{mec}{MEC}{Mobile Edge Cloud}
\newacronym{mi}{MI}{Mutual Information}
\newacronym{mimo}{MIMO}{Multiple Input Multiple Output}
\newacronym{mmwave}{mmWave}{millimeter wave}
\newacronym{mr}{MR}{Maximum Rate}
\newacronym{mss}{MSS}{Maximum Segment Size}
\newacronym{mtd}{MTD}{Machine-Type Device}
\newacronym{mtu}{MTU}{Maximum Transmission Unit}
\newacronym{nfv}{NFV}{Network Function Virtualization}
\newacronym{nlos}{NLOS}{Non Line of Sight}
\newacronym{nr}{NR}{New Radio}
\newacronym{ofdm}{OFDM}{Orthogonal Frequency Division Multiplexing}
\newacronym{pdcch}{PDCCH}{Physical Downlonk Control Channel}
\newacronym{pdcp}{PDCP}{Packet Data Convergence Protocol}
\newacronym{pdsch}{PDSCH}{Physical Downlink Shared Channel}
\newacronym{pdu}{PDU}{Packet Data Unit}
\newacronym{pf}{PF}{Proportional Fair}
\newacronym{pgw}{PGW}{Packet Gateway}
\newacronym{phy}{PHY}{Physical}
\newacronym{pbch}{PBCH}{Physical Broadcast Channel}
\newacronym[plural=\gls{mme}s,firstplural=Mobility Management Entities (MMEs)]{mme}{MME}{Mobility Management Entity}
\newacronym{prb}{PRB}{Physical Resource Block}
\newacronym{pss}{PSS}{Primary Synchronization Signal}
\newacronym{pucch}{PUCCH}{Physical Uplink Control Channel}
\newacronym{pusch}{PUSCH}{Physical Uplink Shared Channel}
\newacronym{rach}{RACH}{Random Access Channel}
\newacronym{ran}{RAN}{Radio Access Network}
\newacronym{red}{RED}{Robotics Emergency Deployment}
\newacronym{rf}{RF}{Radio Frequency}
\newacronym{rlc}{RLC}{Radio Link Control}
\newacronym{rlf}{RLF}{Radio Link Failure}
\newacronym{rrc}{RRC}{Radio Resource Control}
\newacronym{rrm}{RRM}{Radio Resource Management}
\newacronym{rr}{RR}{Round Robin}
\newacronym{rs}{RS}{Remote Server}
\newacronym{rsrp}{RSRP}{Reference Signal Received Power}
\newacronym{rss}{RSS}{Received Signal Strength}
\newacronym{rtt}{RTT}{Round Trip Time}
\newacronym{rw}{RW}{Receive Window}
\newacronym{rx}{RX}{Receiver}
\newacronym{sa}{SA}{standalone}
\newacronym{sack}{SACK}{Selective Acknowledgment}
\newacronym{sap}{SAP}{Service Access Point}
\newacronym{sch}{SCH}{Secondary Cell Handover}
\newacronym{scoot}{SCOOT}{Split Cycle Offset Optimization Technique}
\newacronym{sdma}{SDMA}{Spatial Division Multiple Access}
\newacronym{sinr}{SINR}{Signal to Interference plus Noise Ratio}
\newacronym{sm}{SM}{Saturation Mode}
\newacronym{snr}{SNR}{Signal to Noise Ratio}
\newacronym{son}{SON}{Self-Organizing Network}
\newacronym{ss}{SS}{Synchronization Signal}
\newacronym{srs}{SRS}{Sounding Reference Signal}
\newacronym{sss}{SSS}{Secondary Synchronization Signal}
\newacronym{tb}{TB}{Transport Block}
\newacronym{tcp}{TCP}{Transmission Control Protocol}
\newacronym{tdd}{TDD}{Time Division Duplexing}
\newacronym{tdma}{TDMA}{Time Division Multiple Access}
\newacronym{tfl}{TfL}{Transport for London}
\newacronym{tm}{TM}{Transparent Mode}
\newacronym{trp}{TRP}{Transmitter Receiver Pair}
\newacronym{tti}{TTI}{Transmission Time Interval}
\newacronym{ttt}{TTT}{Time-to-Trigger}
\newacronym{tx}{TX}{Transmitter}
\newacronym{ue}{UE}{User Equipment}
\newacronym{ul}{UL}{Uplink}
\newacronym{uml}{UML}{Unified Modeling Language}
\newacronym{um}{UM}{Unacknowledged Mode}
\newacronym{utc}{UTC}{Urban Traffic Control}
\newacronym{vm}{VM}{Virtual Machine}
\newacronym{rsrq}{RSRQ}{Reference Signal Received Quality}
\newacronym{rssi}{RSSI}{Received Signal Strength Indicator}
\newacronym{crs}{CRS}{Cell Reference Signal}
\newacronym{comp}{CoMP}{Coordinated Multi-Point}
\newacronym{cran}{C-RAN}{Cloud \acrlong{ran}}
\newacronym{ca}{CA}{Carrier Aggregation}
\newacronym{cco}{CC}{Carrier Component}
\newacronym{nsa}{NSA}{Non Stand Alone}
\newacronym{embb}{eMBB}{Enhanced Mobility Broadband}
\newacronym{bsr}{BSR}{Buffer Status Report}
\newacronym{srb}{SRB}{Service Radio Bearer}
\newacronym{scm}{SCM}{Spatial Channel Model}
\newacronym{sctp}{SCTP}{Stream Control Transmission Protocol}
\newacronym{mptcp}{MPTCP}{Multi-path TCP}
\newacronym{ietf}{IETF}{Internet Engineering Task Force}
\newacronym{os}{OS}{Operating System}
\newacronym{tls}{TLS}{Transport Layer Security}
\newacronym{rfc}{RFC}{Request for Comments}
\newacronym{http}{HTTP}{HyperText Transfer Protocol}
\newacronym{nat}{NAT}{Network Address Translation}
\newacronym{api}{API}{Application Programming Interface}
\newacronym{rto}{RTO}{Retransmission Timeout}
\newacronym{psc}{PSC}{Public Safety Communication}
\newacronym{rpgm}{RPGM}{Reference Point Group Mobility}
\newacronym{ic}{IC}{Incident Command}
\newacronym{rsu}{RSU}{Road Side Unit}
\newacronym{uav}{UAV}{unmanned aerial vehicle}
\newacronym{usv}{USV}{Unmanned Surface Vehicle}
\newacronym{uas}{UAS}{Unmanned Aerial System}
\newacronym{iab}{IAB}{Integrated Access and Backhaul}
\newacronym{qoe}{QoE}{Quality of Experience}
\newacronym{ssim}{SSIM}{Structural Similarity Index}
\newacronym{psnr}{PSNR}{Peak Signal to Noise Ratio}
\newacronym{bs}{BS}{Base Station}
\newacronym{mu}{MU}{Multiple User}
\newacronym{ag}{AG}{Air-to-Ground}
\newacronym{af}{AF}{Array Factor}
\newacronym{ula}{ULA}{Uniform Linear Array}
\newacronym{upa}{UPA}{Uniform Planar Array}
\newacronym{lcs}{LCS}{Local Coordinate System}
\newacronym{psd}{PSD}{Power Spectral Density}
\newacronym{vq}{VQ}{vector quantization}
\newacronym{a2g}{A2G}{air-to-ground}
\newacronym{em}{EM}{electromagnetic}
\newacronym{vae}{VAE}{variational autoencoder}
\newacronym{hls}{HLS}{High-Level Synthesis}
\newacronym{sram}{SRAM}{Static Random Access Memory}
\newacronym{fir}{FIR}{Finite Impulse Response}
\newacronym{pe}{PE}{Processing Element}
\newacronym{isac}{ISAC}{Integrated Sensing and Communications}

\begin{document}

\title{A Spatial Array for Spectrally Agile Wireless Processing}

\author{
\IEEEauthorblockN{
Ali Rasteh,
Andrew Hennessee,
Ishaan Shivhare,
Siddharth Garg,
Sundeep Rangan,
Brandon Reagen
}

\thanks{Authors are with Tandon School of Engineering, New York University, Brooklyn, NY, USA.
Rangan and Rasteh are supported in part by NSF grants 1952180, 2133662, 2236097, 2148293, and 1925079, the DARPA Prowess program, the NTIA and industrial affiliates of NYU Wireless.
}
}

\maketitle


\begin{abstract}
Massive MIMO is a cornerstone of next-generation wireless communication, offering significant gains in capacity, reliability, and energy efficiency. However, to meet emerging demands such as high-frequency operation, wide bandwidths, co-existence, integrated sensing, and resilience to dynamic interference, future systems must exhibit both scalability and spectral agility. These requirements place increasing pressure on the underlying processing hardware to be both efficient and reconfigurable. This paper proposes a custom-designed spatial array architecture that serves as a reconfigurable, general-purpose core optimized for a class of wireless kernels that commonly arise in diverse
communications and sensing tasks.
The proposed spatial array is evaluated against specialized cores for each kernel using  High-Level Synthesis (HLS). Both the reconfigurable and specialized designs are synthesized in a \SI{32}{nm} process to assess latency, throughput, area, and power in realistic processes. The results identify conditions under which general-purpose systolic architectures can approach the efficiency of specialized cores, thereby paving the way toward more scalable and agile systems.
\end{abstract}


\section{Introduction}
Massive \gls{mimo} \cite{marzetta2002capacity},
where the base stations use a large number of antenna elements and streams,
was one of the most critical technologies for increasing capacity in 5G systems \cite{larsson2014massive}.
However, the hardware to support the next generation system faces at least two significant challenges:

\medskip

\noindent
\underline{Scalability}:
There is now considerable interest in expanding the \gls{mimo} antenna dimensions to unprecedented sizes \cite{sanguinetti2019toward}.  For example, the simulation study \cite{nokia2025massiveMIMO}
shows that \gls{mimo} systems with 1024
antenna elements (at least five times greater than current commercial base stations) can increase the spectral efficiency by at least four fold.  Such massive \gls{mimo} systems, sometimes called \emph{extreme \gls{mimo}} \cite{wesemann2023energy}, are particularly valuable in the emerging upper mid-band 
\cite{kang2024cellular}.
Moreover, in addition to the capacity gains, high dimensional arrays can provide significant benefits for
interference cancellation \cite{jia2025joint} and the development of wide bandwidth systems \cite{akrout2023bandwidth}.

\noindent
\underline{Spectral agility}:  Future systems will likely 
perform a much wider and more dynamic range of spectral tasks
than traditional cellular communications transceivers.
For example, dynamic spectrum sharing -- a key feature in 5G \cite{ahmad20205g} -- requires new
hardware for interference sensing and nulling,
as well as highly variable bandwidths \cite{karkhaneh2024implementation}.
Spectrum sharing is particularly vital for emerging non-terrestrial networks (NTN) 
\cite{kang2024cellular,jia2025joint}.
In addition, \gls{isac} requires hardware for RADAR, localization,
and RF imaging \cite{wu2025integrated}.

At root, these two demands are in tension.
Spectral agility demands reconfigurability
in the hardware, which generally comes
at a cost of the processing capability
for a fixed area and power.
However, systems will need to significantly increase processing capabilities to meet the scaling requirements in terms of the number of antennas and bandwidths
for  next-generation systems.
A fundamental question is how to develop efficient hardware that can be both reconfigurable and efficient in processing.


\subsection*{Contributions}
To address these challenges, our contributions are as follows:
\begin{itemize}
    \item \emph{Novel spatial array}:  
    We present a novel compute unit that we call a \emph{spatial array}, which can provide a building block for a wide range of tasks.
    The spatial array can be seen as a flexible systolic array, which is widely-used in machine learning \cite{xu2023survey}—and has also been proposed in domain-adaptive processors for wireless communication~\cite{weng2020hybrid,chen2024dap}.
    While classical systolic arrays are optimized for matrix multiplication, the proposed  architecture is designed to accommodate a diverse set of critical kernels within the wireless domain, including \gls{fir} filtering, convolutions, matrix-matrix and matrix-vector multiplication, and outer-product computations, among others.

    \item \emph{Workload enumeration}:
    We enumerate a number of core kernels that are required 
    for diverse tasks, including filtering, channelization, equalization, and MIMO processing for communications, as well as outer products and matched filters for spectrum sensing.
    
    \item \emph{Comparison to specialized hardware}:  For each kernel, we compare the proposed spatial array with custom hardware developed by \gls{hls}~\cite{coussy2010high}. Both the proposed reconfigurable spatial array and the custom designs are synthesized in \SI{32}{nm} technology node to assess latency, throughput, area, and power consumption.  Our results show that
    the reconfigurable engine provides performance close to that of specialized hardware for each task while enabling reconfigurability.
\end{itemize}

\begin{figure*}[]
  \centering
  \includegraphics[width=0.8\textwidth]{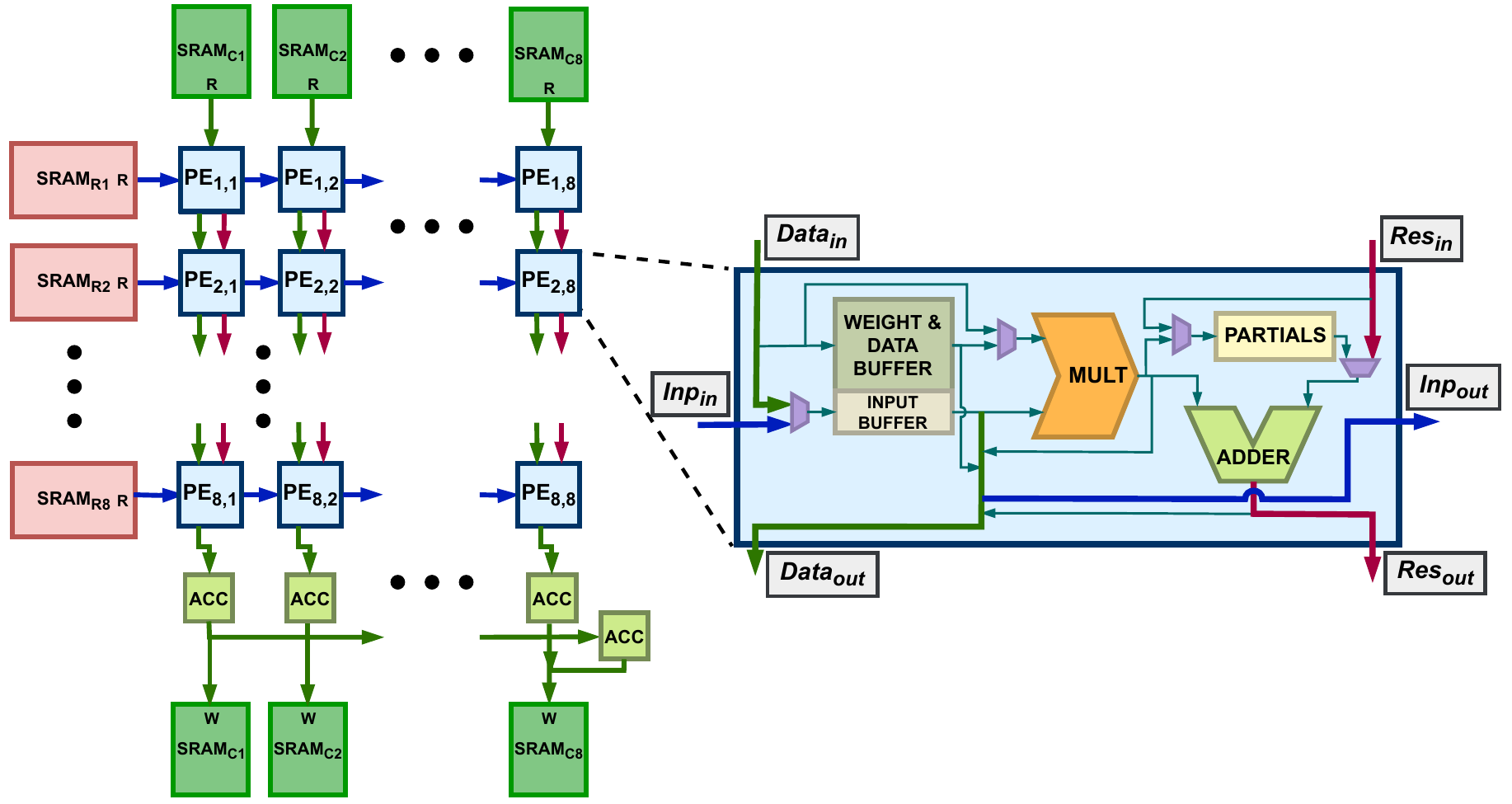}
    \caption{The architectural design of the proposed spatial array is depicted. The illustration on the left presents our $8 \times 8$ array, demonstrating SRAM connections accessible from the upper, lower, and lateral sides of the array, facilitating adaptable data injection under various conditions. The illustration on the right provides an in-depth representation of the architecture of our processing element.
  \label{fig:spatial_array}}
\end{figure*}

\section{Reconfigurable Spatial Array Architecture and Data Flow}
Figure~\ref{fig:spatial_array} shows the layout and design of the proposed spatial array.
The array is structured as a grid of \glspl{pe} and primarily operates using a weight-stationary data flow. In this scheme, one operand—typically the weight—is stored within the array, while the other operand(s) are streamed in. Each \gls{pe} includes a multiplier, an adder, and a buffer of registers. The \glspl{pe} supports two modes of operation: accumulate mode and element-wise mode. Accumulate mode is well-suited for kernels such as convolution, where the partial products of each convolution are summed to produce the output. Element-wise mode enables support for operations like vector magnitude squared, where individual output elements are routed using the buffer.
The buffer also allows for the storage of multiple weight tiles, reducing memory reads and improving input reuse for compute-intensive kernels such as matrix-matrix multiplication. Each row and column of the array is connected to a dedicated \gls{sram} bank for data input: weights are typically streamed from the top (column-wise \gls{sram}), while inputs enter from the left (row-wise \gls{sram}). Some kernels require both weights and inputs to be delivered from the top \gls{sram}. Outputs are produced at the bottom of the array, where a row of accumulators handles the further accumulation of partial results before writing them to memory.
In our setup, we assume that the top \gls{sram} supports two reads per cycle, while the left \gls{sram} supports one read per cycle.

\section{Performance Evaluation and Comparative Analysis with HLS}

To compare the performance of the proposed reconfigurable spatial array with \gls{hls} implementations, we selected a set of highly utilized kernels common in wireless baseband processing. The evaluation set includes matrix-vector multiplication, matrix-matrix multiplication, \gls{fir} filtering, matched filtering, vector magnitude squaring, and outer product. For the spatial array, we developed specific hardware mappings for these kernels and measured the key performance metrics: latency, throughput, utilization, area, and power consumption. Conversely, we developed equivalent \gls{hls} implementations for each kernel to measure their corresponding latency, throughput, area, and power. Performance was rigorously assessed under various conditions, considering real and complex data types as well as different sizes for input and weight vectors. All simulations were performed using the SAED \SI{32}{nm} Low-Voltage Threshold (LVT) technology node, Catapult Ultra Synthesis 2023.1\_2, and Synopsys Design Compiler T-2022.03-SP5-5.
In our hardware implementations, particularly within the \gls{hls} framework, we adopt the constraint that the total available \gls{sram} bandwidth is the primary system bottleneck. This bottleneck is directly proportional to the number of available \gls{sram} read/write ports for the on-chip memory bank serving the computational units. Crucially, we assume that the available memory bandwidth is identical for both the \gls{hls} kernels and the spatial array.

\begin{table*}[]
\caption{Key performance metrics (latency, utilization, throughput, and power consumption) for various kernels with different input and weight configurations. 'Lower Bound' represents the theoretical minimum number of cycles required to complete each computation under the given constraints. In evaluating throughput, it is presupposed that each multiplication and addition constitutes an independent operation.
}
\label{tab:performance_sa}
\centering
\footnotesize
\renewcommand{\arraystretch}{1.1}
\begin{tabular}{>{\centering\arraybackslash}m{2.1cm}||>{\centering\arraybackslash}m{1.2cm}|>{\centering\arraybackslash}m{1.1cm}||>{\centering\arraybackslash}m{1.1cm}|>{\centering\arraybackslash}m{1.1cm}||>{\centering\arraybackslash}m{1.0cm}|>{\centering\arraybackslash}m{1.1cm}|>{\centering\arraybackslash}m{1.5cm}|>{\centering\arraybackslash}m{1.4cm}|>{\centering\arraybackslash}m{1.0cm}}
\hline\hline
\multirow{2}{*}{\textbf{Kernel}}
  & \multicolumn{2}{c||}{\textbf{Inputs}}
  & \multicolumn{2}{c||}{\textbf{Weights}}
  & \multirow{2}{*}{\shortstack{\textbf{Latency} \\ \textbf{(Cycles)}}}
  & \multirow{2}{*}{\shortstack{\textbf{Utilization} \\ \textbf{(\%)}}}
  & \multirow{2}{*}{\shortstack{\textbf{Lower Bound} \\ \textbf{(Cycles)}}}
  & \multirow{2}{*}{\shortstack{\textbf{Throughput} \\ \textbf{(GOPS)}}}
  & \multirow{2}{*}{\shortstack{\textbf{Power} \\ \textbf{($m W$)}}}
  \tabularnewline
\cline{2-5}
  & \textbf{Size} & \textbf{Type}
  & \textbf{Size} & \textbf{Type}
  &  &  & & &
  \tabularnewline
\hline\hline
\multirow{3}{*}{\shortstack{\textbf{Matrix Vector} \\ \textbf{Multiplication}}}
  & (1024,4)  & \multirow{3}{*}{Complex}
  & (4,1)     & \multirow{3}{*}{Complex}
  &  530      & 48.3   & 256 & 61.83 & \num{93.4}
  \tabularnewline
  & (1024,8)  & 
  & (8,1)     & 
  & 1042      & 49.1   & 512 & 62.89 & \num{95.0} 
  \tabularnewline
  & (1024,16) & 
  & (16,1)    & 
  & 2066      & 49.5   & 1024 & 63.44 & \num{95.7} 
  \tabularnewline
\hline
\multirow{6}{*}{\shortstack{\textbf{Matrix} \\ \textbf{Multiplication}}}
  & (1024,4)   & \multirow{6}{*}{Real}
  & (4,8)      & \multirow{6}{*}{Real}
  &  527       & 97.4   & 512 & 124.36 & \num{188.0}
  \tabularnewline
  & (1024,4)   & 
  & (4,16)     & 
  & 1039       & 98.5   & 1024 & 126.15 & \num{191.0}
  \tabularnewline
  & (1024,8)   & 
  & (8,8)      & 
  & 1039       & 98.5   & 1024 & 126.15 & \num{191.0}
  \tabularnewline
  & (1024,8)   & 
  & (8,16)     & 
  & 2063       & 99.2   & 2048 & 127.07 & \num{192.0}
  \tabularnewline
  & (1024,16)  & 
  & (16,8)     & 
  & 2063       & 99.2   & 2048 & 127.07 & \num{192.0}
  \tabularnewline
  & (1024,16)  & 
  & (16,16)    & 
  & 4119       & 99.44  & 4096 & 127.28 & \num{192.0}
  \tabularnewline
\hline
\multirow{2}{*}{\textbf{FIR Filter}}
  & (1024,1)   & Real
  & (32,1)     & Real
  &  464       & 48.27  & 224 & 61.79 & \num{93.4}
  \tabularnewline
  & (1024,1)   & Complex
  & (32,1)     & Complex
  &  912       & 98.2   & 894 & 125.47 & \num{190.0}
  \tabularnewline
\hline
\multirow{2}{*}{\textbf{Matched Filter}}
  & 1×(1024,1) & \multirow{2}{*}{Complex}
  & (32,1)   & \multirow{2}{*}{Complex}
  &  2,232 & 91.8 & 2050 & 117.56 & \num{48.4}
  \tabularnewline
  & 8×(1024,1) & 
  & (32,1)   & 
  &  18,180 & 90.20 & 16400 & 115.46 & \num{180.0}
  \tabularnewline
\hline
\multirow{2}{*}{\shortstack{\textbf{Vector Magnitude} \\ \textbf{Squared}}}
  & (512,1)    & \multirow{2}{*}{Complex}
  & (512,1)    & \multirow{2}{*}{Complex}
  &  64       & 50   & 32 & 64 & \num{48.6}
  \tabularnewline
  & (1024,1)   & 
  & (1024,1)   & 
  &  128       & 50   & 64 & 64 & \num{48.6}
  \tabularnewline
\hline
\multirow{5}{*}{\shortstack{\textbf{Outer} \\ \textbf{Product}}}
  & (1024,8)   & 
  & (1024,8)   & 
  &  4132      & 99.12  & 4096 & 126.88 & \num{192.0}
  \tabularnewline
  & (1024,32)  & \multirow{3}{*}{Complex}
  & (1024,32)  & \multirow{3}{*}{Complex}
  &  65,572    & 99.94  & 65,536 & 127.93 & \num{193.0}
  \tabularnewline
  & (1024,64)  & 
  & (1024,64)  & 
  & 262,180 & 99.98 & 262,144 & 127.98 & \num{193.0}
  \tabularnewline
  & (1024,128) & 
  & (1024,128) & 
  & 1.05M & 99.98 & 1.048M & 127.98 & \num{193.0}
  \tabularnewline
  & (1024,512) & 
  & (1024,512) & 
  & 16.78M & 99.99 & 16.78M & 127.99 & \num{193.0}
  \tabularnewline
\hline\hline
\end{tabular}
\end{table*}

\subsection{Performance Analysis of the Spatial Array}

Table~\ref{tab:performance_sa} provides an overview of the initial performance metrics of the spatial array architecture, specifically detailing latency, utilization, throughput, and power consumption. The measured area of the synthesized spatial array is \(1.014\,\text{mm}^2\). The Lower-Bound column represents the theoretical minimum number of cycles required for an operation to complete. This is determined by the total number of arithmetic operations (e.g., multiplications) necessary for the computation, divided by the number of arithmetic units (multipliers) present within the array.
Kernels exhibiting suboptimal performance often lack input data reuse, leading to a performance bottleneck primarily constrained by the available \gls{sram} bandwidth. This indicates that the time spent waiting for new data from memory exceeds the time spent on computation. The \gls{fir} filter demonstrates comparatively lower performance when processing real-valued data compared to complex-valued data. This counter-intuitive result is due to the inherent structure of the spatial array's \gls{pe} and the timing of the operations. The increased duration required for the completion of a complex multiplication (which typically involves four real multiplications and two additions) provides a larger temporal margin for input shifting and subsequent data reuse across the array, thereby mitigating the memory bandwidth bottleneck for the complex-valued case.

\begin{table*}[]
\caption{Key performance metrics (latency, throughput, area, and power consumption) for various wireless kernels implemented using \gls{hls}, evaluated across distinct input and weight configurations.
}
\label{tab:performance_hls}
\centering
\footnotesize
\renewcommand{\arraystretch}{1.1}
\begin{tabular}{>{\centering\arraybackslash}m{2.1cm}||>{\centering\arraybackslash}m{1.2cm}|>{\centering\arraybackslash}m{1.1cm}||>{\centering\arraybackslash}m{1.1cm}|>{\centering\arraybackslash}m{1.1cm}||>{\centering\arraybackslash}m{1.0cm}|>{\centering\arraybackslash}m{1.4cm}|>{\centering\arraybackslash}m{1.0cm}|>{\centering\arraybackslash}m{1.0cm}}
\hline\hline
\multirow{2}{*}{\textbf{Kernel}}
  & \multicolumn{2}{c||}{\textbf{Inputs}}
  & \multicolumn{2}{c||}{\textbf{Weights}}
  & \multirow{2}{*}{\shortstack{\textbf{Latency} \\ \textbf{(Cycles)}}}
  & \multirow{2}{*}{\shortstack{\textbf{Throughput} \\ \textbf{(GOPS)}}}
  & \multirow{2}{*}{\shortstack{\textbf{Area} \\ \textbf{($m m^2$)}}}
  & \multirow{2}{*}{\shortstack{\textbf{Power} \\ \textbf{($m W$)}}}
  \tabularnewline
\cline{2-5}
  & \textbf{Size} & \textbf{Type}
  & \textbf{Size} & \textbf{Type}
  &  &  &
  \tabularnewline
\hline\hline
\multirow{3}{*}{\shortstack{\textbf{Matrix Vector} \\ \textbf{Multiplication}}}
  & (1024,4)  & \multirow{3}{*}{Complex}
  & (4,1)     & \multirow{3}{*}{Complex}
  & 540 & 60.68 & 0.3528 & \num{91.0}
  \tabularnewline
  & (1024,8)  & 
  & (8,1)     & 
  & 1072 & 61.13 & 0.4637 & \num{108.0}
  \tabularnewline
  & (1024,16) & 
  & (16,1)    & 
  & 2144 & 61.13 & 0.4637 & \num{108.0}
  \tabularnewline
\hline
\multirow{6}{*}{\shortstack{\textbf{Matrix} \\ \textbf{Multiplication}}}
  & (1024,4)   & \multirow{6}{*}{Real}
  & (4,8)      & \multirow{6}{*}{Real}
  & 2070 & 31.66 & 0.2018 & \num{58.2}
  \tabularnewline
  & (1024,4)   & 
  & (4,16)     & 
  & 4118 & 31.83 & 0.2025 & \num{58.3}
  \tabularnewline
  & (1024,8)   & 
  & (8,8)      & 
  & 4138 & 31.68 & 0.2641 & \num{74.0}
  \tabularnewline
  & (1024,8)   & 
  & (8,16)     & 
  & 8234 & 31.84 & 0.2650 & \num{74.6}
  \tabularnewline
  & (1024,16)  & 
  & (16,8)     & 
  & 8274 & 31.68 & 0.4029 & \num{107.0}
  \tabularnewline
  & (1024,16)  & 
  & (16,16)    & 
  & 16,466 & 31.84 & 0.4054 & \num{108.0}
  \tabularnewline
\hline
\multirow{2}{*}{\textbf{FIR Filter}}
  & (1024,1)   & Real
  & (32,1)     & Real
  & 2150 & 29.53 & 0.4004 & 107.9
  \tabularnewline
  & (1024,1)   & Complex
  & (32,1)     & Complex
  &  4160 & 61.05 & 0.4641 & 107.2
  \tabularnewline
\hline
\multirow{2}{*}{\textbf{Matched Filter}}
  & 1×(1024,1) & \multirow{2}{*}{Complex}
  & (32
  ,1)   & \multirow{2}{*}{Complex}
  & 4160 & 61.05 & 0.4641 & 107.2
  \tabularnewline
  & 8×(1024,1) & 
  & (32,1)   & 
  & 33,280 & 61.05 & 0.4641 & 107.2
  \tabularnewline
\hline
\multirow{2}{*}{\shortstack{\textbf{Vector Magnitude} \\ \textbf{Squared}}}
  & (512,1)    & \multirow{2}{*}{Complex}
  & (512,1)    & \multirow{2}{*}{Complex}
  & 135 & 30.34 & 0.1164 & \num{28.0}
  \tabularnewline
  & (1024,1)   & 
  & (1024,1)   & 
  & 263 & 31.15 & 0.1141 & \num{28.2}
  \tabularnewline
\hline
\multirow{5}{*}{\shortstack{\textbf{Outer} \\ \textbf{Product}}}
  & (1024,8)   & \multirow{5}{*}{Complex}
  & (1024,8)   & \multirow{5}{*}{Complex}
  & 8204 & 63.91 & 0.3296 & \num{74.7}
  \tabularnewline
  & (1024,32)  & 
  & (1024,32)  & 
  & 131,264 & 63.91 & 0.3296 & \num{74.7}
  \tabularnewline
  & (1024,64)  &
  & (1024,64)  &
  & 525,056 & 63.91 & 0.3296 & \num{74.7}
  \tabularnewline
  & (1024,128) & 
  & (1024,128) & 
  & 2.10M & 63.91 & 0.3296 & \num{74.7}
  \tabularnewline
  & (1024,512) & 
  & (1024,512) & 
  & 33.60M & 63.91 & 0.3296 & \num{74.7}
  \tabularnewline
\hline\hline
\end{tabular}
\end{table*}

\subsection{Performance Metrics for Specialized HLS Implementations}

Table~\ref{tab:performance_hls} demonstrates the key performance metrics for the equivalent \gls{hls} implementations. This comparison includes latency, throughput, area, and power consumption, providing a baseline to assess the efficiency and overhead of the customized spatial array architecture against a standard hardware synthesis flow for the same kernels and input/weight characteristics.

\begin{figure}[htbp]
    \centering
    
    %
    \begin{subfigure}[b]{0.24\textwidth}
        \centering
        \includegraphics[width=\linewidth]{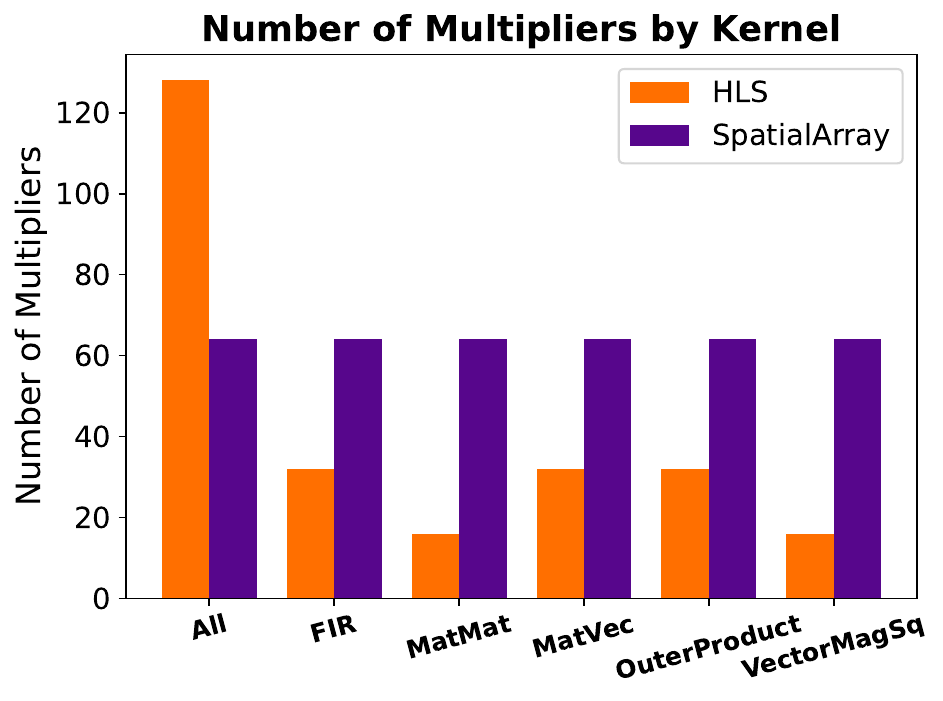}
        \label{fig:bar_multipliers}
    \end{subfigure}
    \begin{subfigure}[b]{0.24\textwidth}
        \centering
        \includegraphics[width=\linewidth]{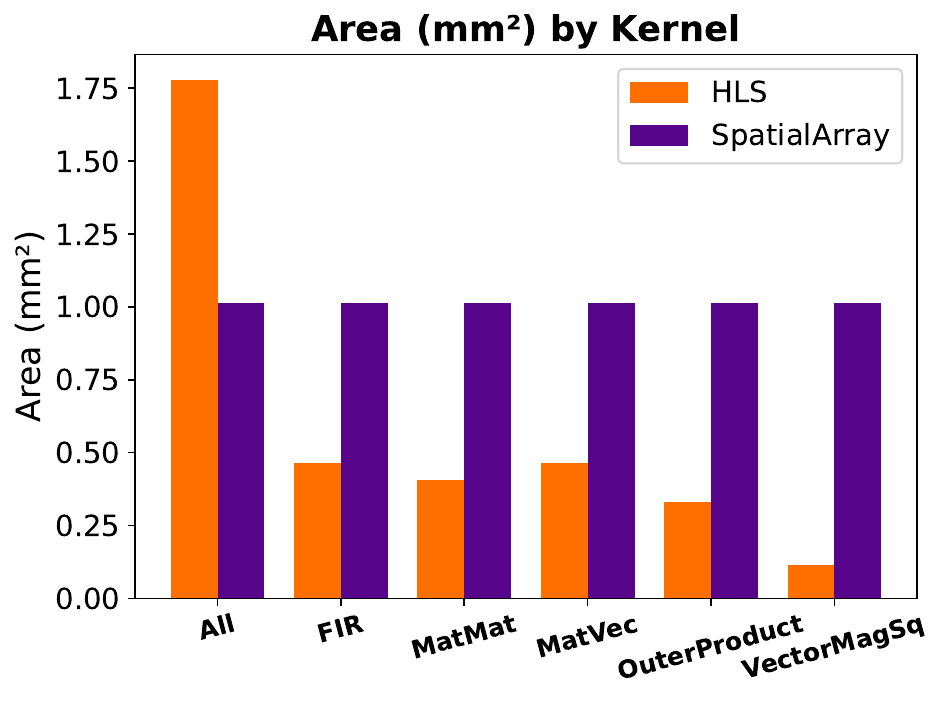}
        \label{fig:bar_area}
    \end{subfigure}
    \begin{subfigure}[b]{0.24\textwidth}
        \centering
        \includegraphics[width=\linewidth]{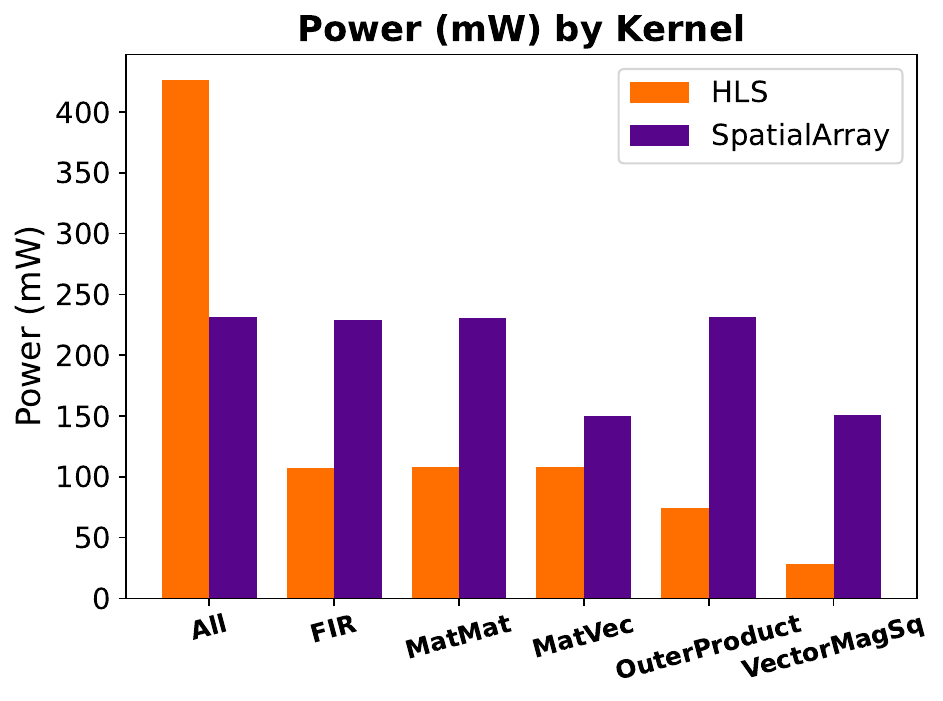}
        \label{fig:bar_power}
    \end{subfigure}

    \vspace{-1em}

    \caption{Number of multipliers, Area ($mm^2$), (c) Power consumption ($mW$) for each kernel in the proposed Spatial Array compared with the \gls{hls} implementations.}
    \label{fig:bar_1}
\end{figure}

\begin{figure}[htbp]
    \centering
    
    \begin{subfigure}[b]{0.24\textwidth}
        \centering
        \includegraphics[width=\linewidth]{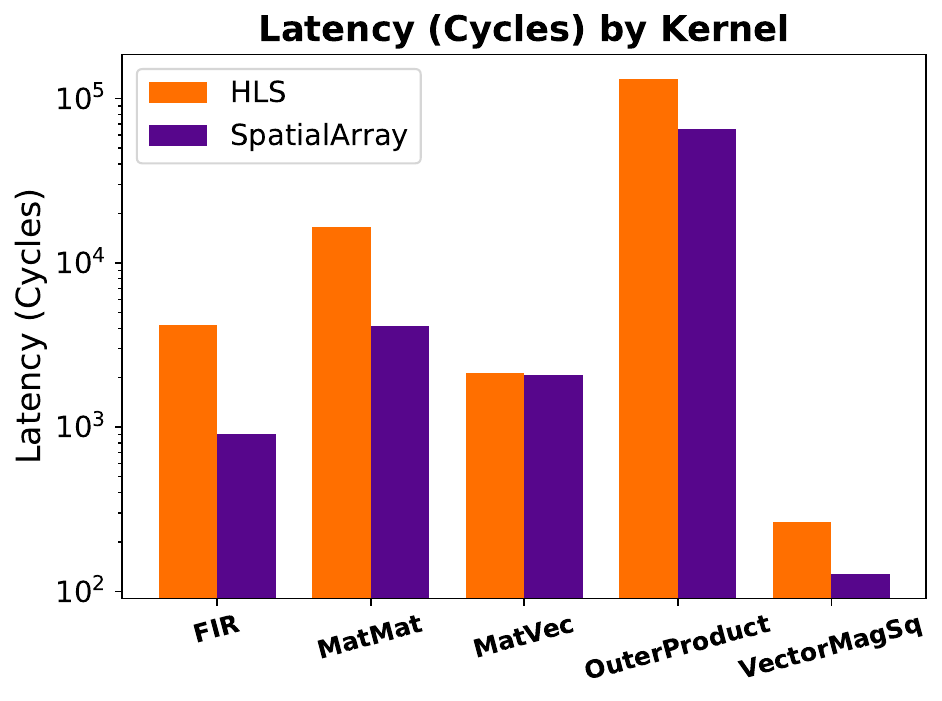}
        \label{fig:bar_latency}
    \end{subfigure}
    \begin{subfigure}[b]{0.24\textwidth}
        \centering
        \includegraphics[width=\linewidth]{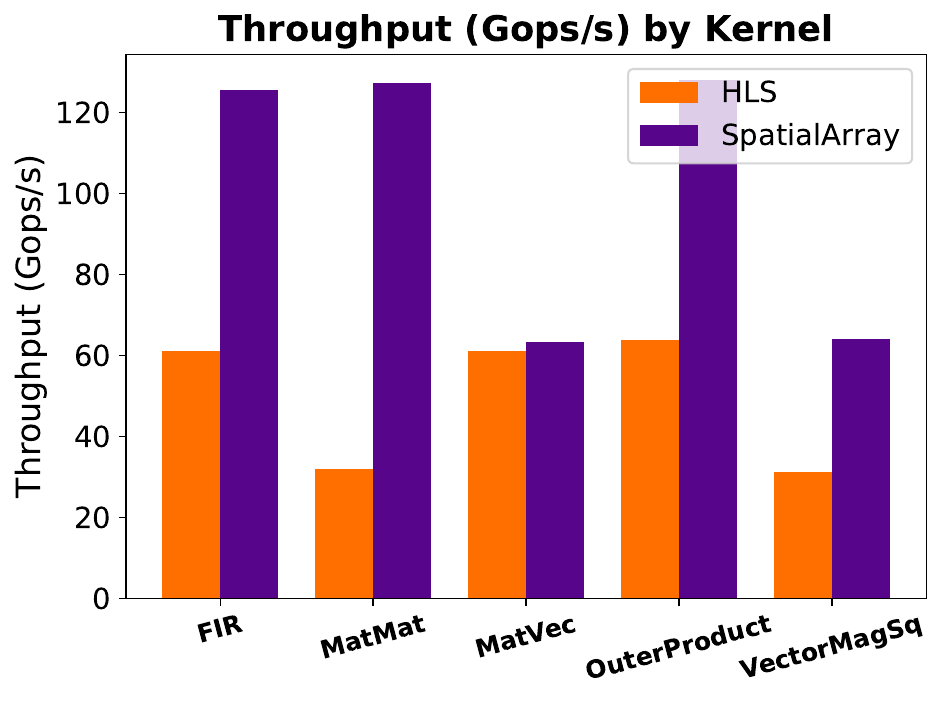}
        \label{fig:bar_throughput}
    \end{subfigure}
    \vspace{-1em}
    \begin{subfigure}[b]{0.24\textwidth}
        \centering
        \includegraphics[width=\linewidth]{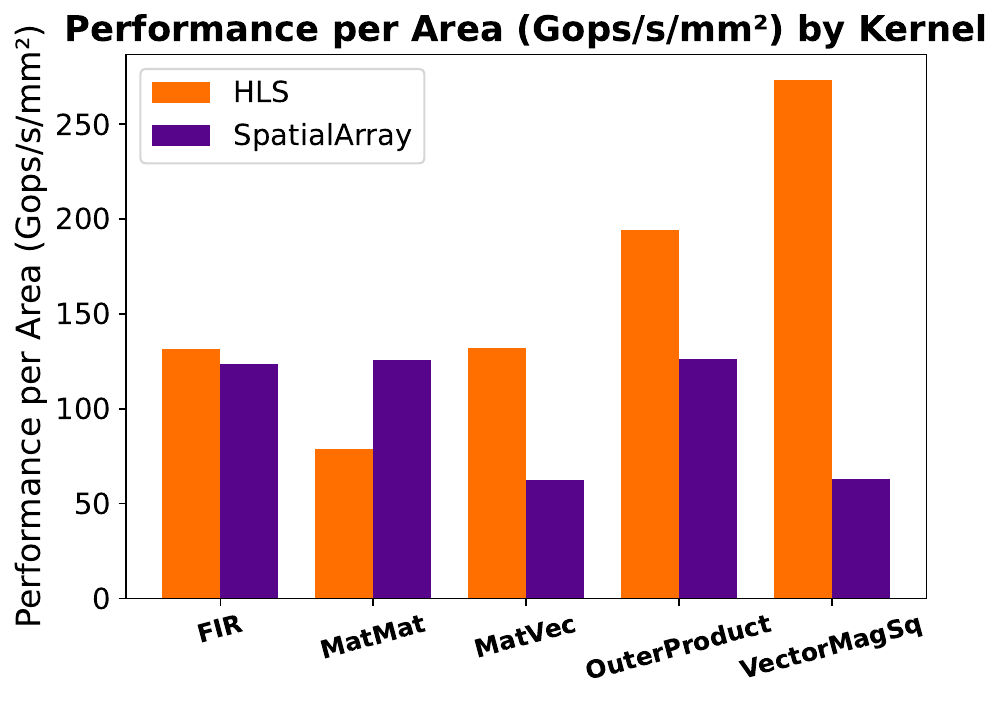}
        \label{fig:bar_perf_per_area}
    \end{subfigure}
    \begin{subfigure}[b]{0.24\textwidth}
        \centering
        \includegraphics[width=\linewidth]{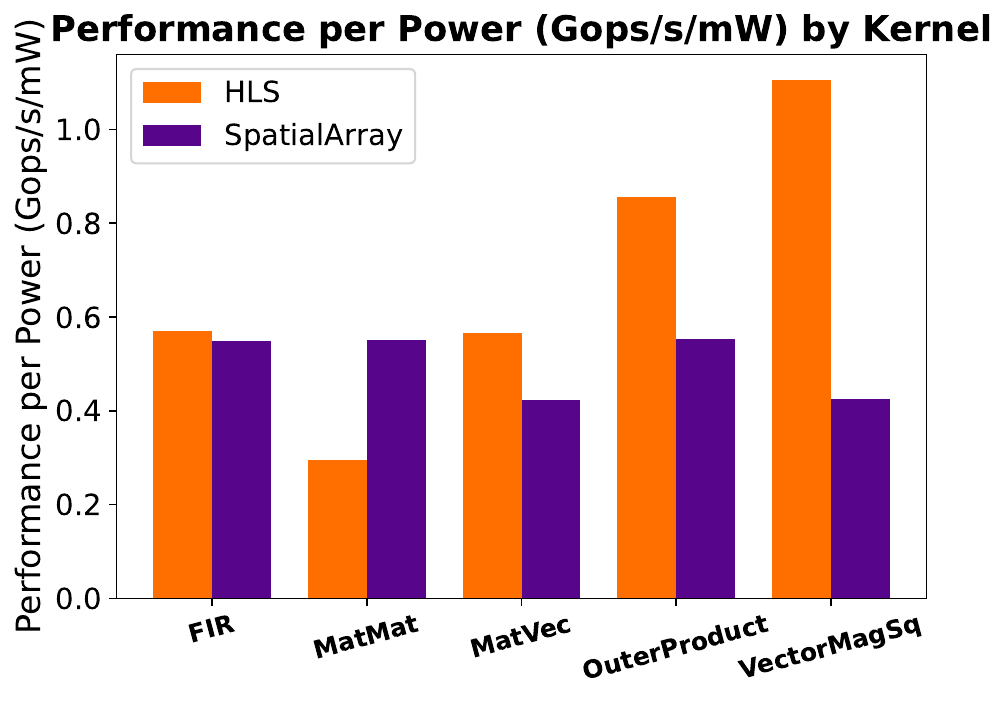}
        \label{fig:bar_perf_per_power}
    \end{subfigure}
    %

    \caption{Latency, Throughput, Performance per area ($Gops/mm^2$), Performance per power ($Gops/mW$) for each kernel in the proposed Spatial Array compared with the \gls{hls} implementations.}
    \label{fig:bar_2}
\end{figure}

\begin{figure*}[htbp]
    \centering
    
    \begin{subfigure}[b]{0.19\textwidth}
        \centering
        \includegraphics[width=\linewidth]{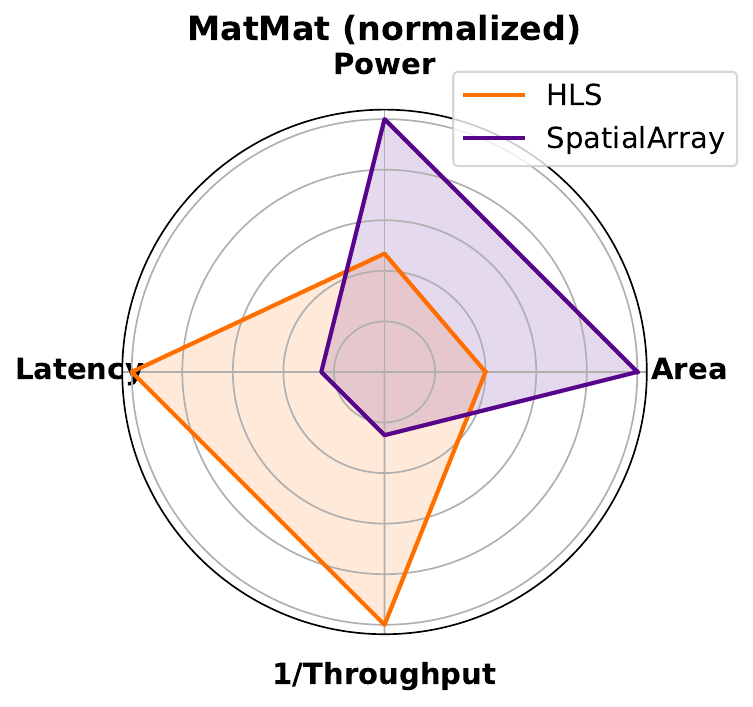}
        \label{fig:radar_MatMat}
    \end{subfigure}
    \begin{subfigure}[b]{0.19\textwidth}
        \centering
        \includegraphics[width=\linewidth]{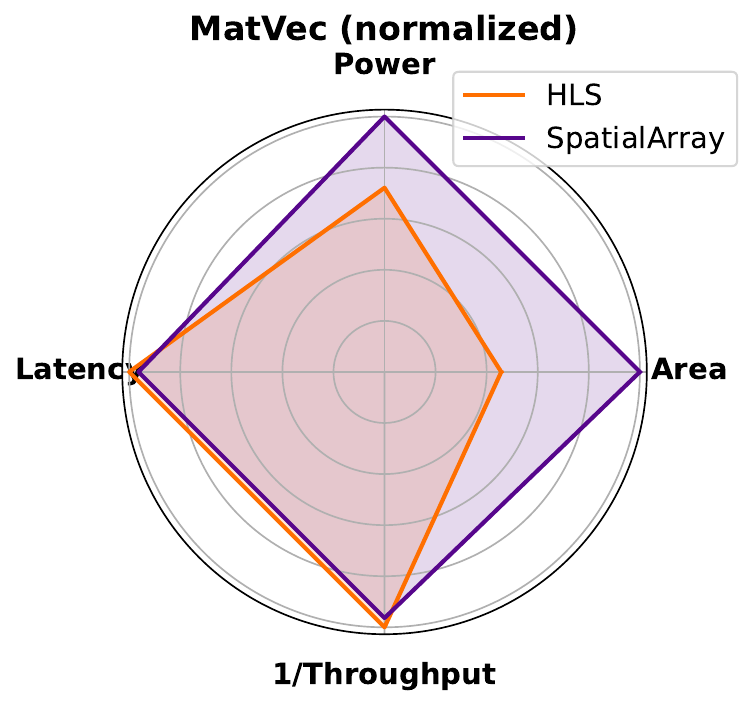}
        \label{fig:radar_MatVec}
    \end{subfigure}
    \begin{subfigure}[b]{0.19\textwidth}
        \centering
        \includegraphics[width=\linewidth]{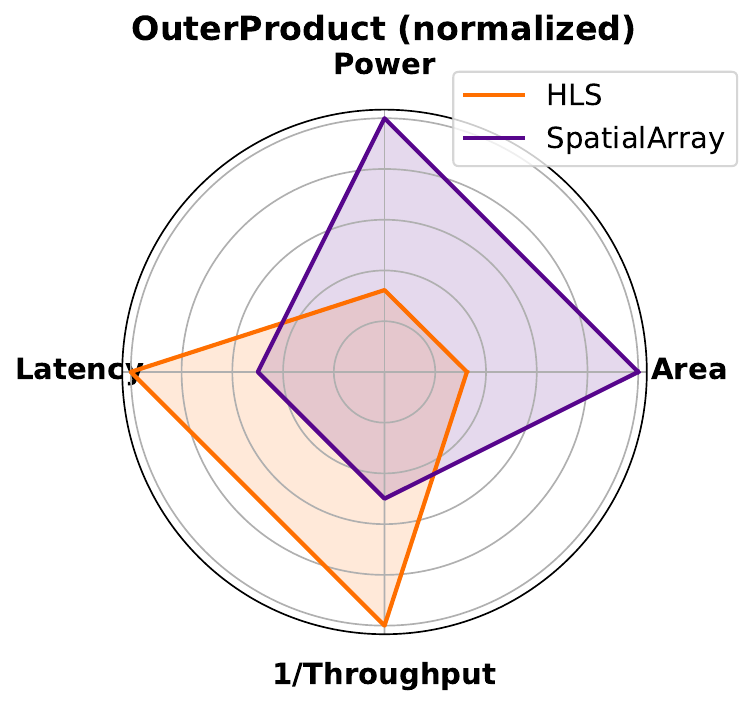}
        \label{fig:radar_OuterProduct}
    \end{subfigure}
    \begin{subfigure}[b]{0.19\textwidth}
        \centering
        \includegraphics[width=\linewidth]{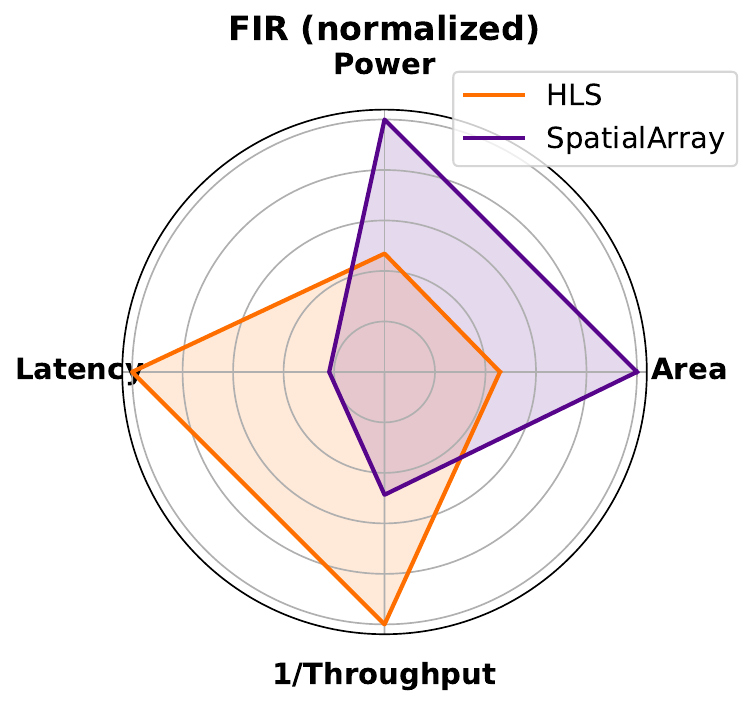}
        \label{fig:radar_FIR}
    \end{subfigure}
    \begin{subfigure}[b]{0.19\textwidth}
        \centering
        \includegraphics[width=\linewidth]{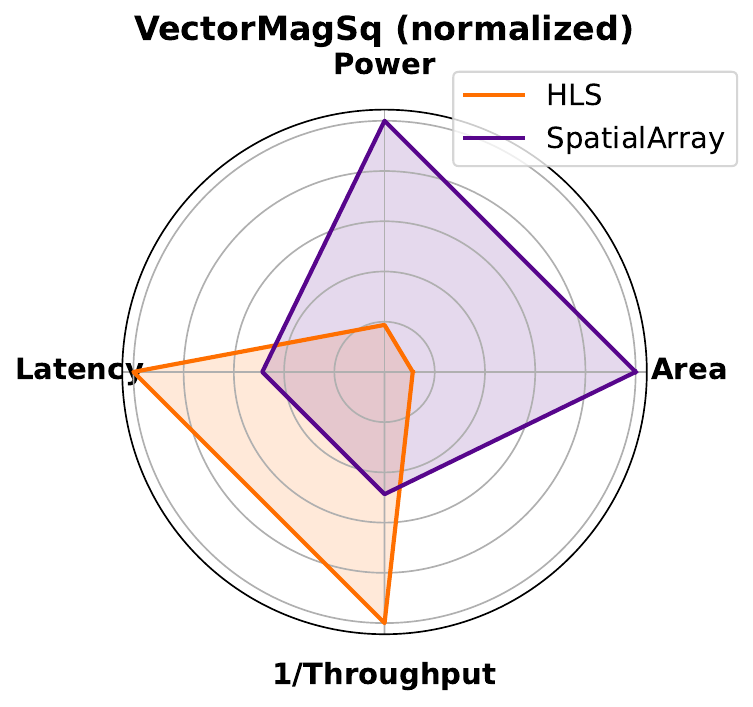}
        \label{fig:radar_VectorMagSq}
    \end{subfigure}

    \vspace{-1em}

    \caption{Radar charts illustrating normalized latency, inverse throughput, area, and power consumption for the proposed spatial array compared with \gls{hls} implementations for five benchmark kernels from left to right: matrix–matrix multiplication, matrix–vector multiplication, outer product, FIR filtering, and vector magnitude squared.}
    \label{fig:radar}
\end{figure*}

\begin{figure}[htbp]
    \centering
    
    \begin{subfigure}[b]{0.24\textwidth}
        \centering
        \includegraphics[width=\linewidth]{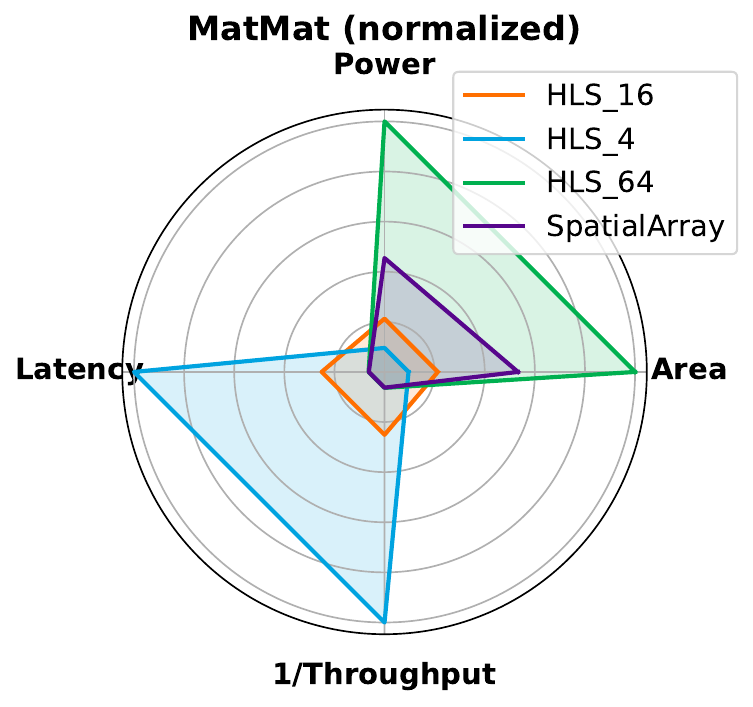}
        \subcaption{}
        \label{fig:radar_MatMat_sweep}
    \end{subfigure}
    \begin{subfigure}[b]{0.24\textwidth}
        \centering
        \includegraphics[width=\linewidth]{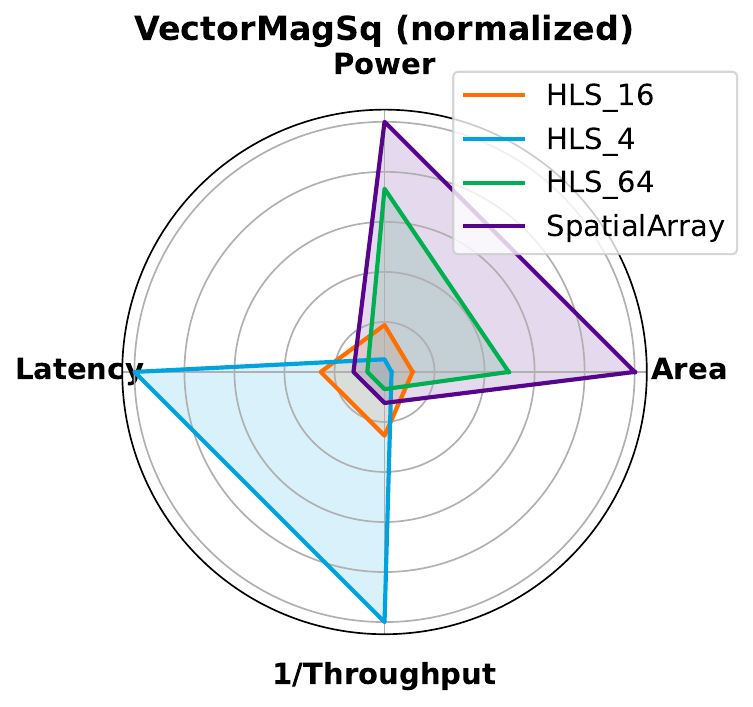}
        \subcaption{}
        \label{fig:radar_VectorMagSq_sweep}
    \end{subfigure}

    \caption{Radar charts depicting normalized latency, inverse throughput, area, and power consumption for the proposed spatial array versus \gls{hls} implementations provisioned with 4, 16, and 64 multipliers across two benchmark kernels from left to right: matrix–matrix multiplication and vector magnitude squared.}
    \label{fig:radar_sweep}
\end{figure}

\subsection{Comparative Visualization of Key Metrics and Trade-offs}

To quantitatively evaluate the distinct aspects of the proposed spatial array and \gls{hls} implementations, we utilize bar charts and radar charts. These visualizations are employed to directly compare the performance characteristics of the respective methods across various operational kernels.
Figures~\ref{fig:bar_1} and \ref{fig:bar_2} present bar charts that compare various metrics across different kernels for the spatial array and \gls{hls} implementations. As previously established, we assume that the available memory bandwidth is identical for both the \gls{hls} kernels and the spatial array. Consequently, the number of utilized multipliers in the \gls{hls} implementations, shown in Figure~\ref{fig:bar_1}, varies across kernels. Conversely, the spatial array, as depicted in Figure~\ref{fig:spatial_array}, consistently utilizes 64 multipliers. Figure~\ref{fig:bar_1} also illustrates the area and power consumption for each method across the different kernels. As anticipated, the highly customized \gls{hls} kernels exhibit reduced area and power consumption compared to the spatial array. A key column in the bar chart is labeled 'All,' which represents a virtual computational unit capable of executing all depicted kernels. For the reconfigurable spatial array, the existing hardware inherently supports all kernels. In contrast, the \gls{hls} implementation requires the aggregation of all individual customized kernels to achieve the same coverage. Therefore, the 'All' column explicitly demonstrates the reconfigurability trade-off between the two methods, highlighting the spatial array's superior performance in this reconfigurable context.

Figure~\ref{fig:bar_2} demonstrates that, with the \gls{hls} kernels utilizing fewer multipliers, the spatial array achieves superior performance in terms of latency and throughput. However, a comparative analysis of throughput per unit of area and throughput per unit of power (shown in Figure~\ref{fig:bar_2}) reveals that the \gls{hls} implementation generally outperforms the spatial array, with the notable exception of matrix multiplication. This exception is attributable to the inherent high efficiency of the spatial array architecture for matrix multiplication, a characteristic further supported by the utilization values presented in Table~\ref{tab:performance_sa}.

Figure~\ref{fig:radar} provides a summary visualization of the preceding data in the form of radar charts, explicitly illustrating the trade-offs among latency, throughput, area, and power consumption across the different kernels and implementation methods.
Finally, Figure~\ref{fig:radar_sweep} provides a performance comparison between the spatial array and the \gls{hls} kernels by sweeping the number of multipliers while maintaining a constant available \gls{sram} bandwidth. This comparison focuses on two extreme kernels in terms of spatial array efficiency: matrix-matrix multiplication and vector magnitude squaring. We evaluate the \gls{hls} implementation across three distinct multiplier configurations (4, 16, and 64). It is observed that for the matrix multiplication kernel, the spatial array is a clear winner across all metrics, even when the \gls{hls} kernel utilizes 64 multipliers (the same as the spatial array), thereby demonstrating highly efficient area and power consumption. Conversely, for the vector magnitude squaring kernel, the spatial array is demonstrably less efficient across all four metrics when compared to the corresponding \gls{hls} implementation.

\section{Conclusion}

This paper presents a comparative analysis between a custom-designed reconfigurable spatial array and specialized High-Level Synthesis (HLS) cores to address the emerging needs of Massive MIMO systems, specifically scalability and spectral agility. By evaluating key wireless kernels—including FIR filtering, matrix multiplication, and outer product—under identical memory bandwidth constraints, we demonstrate distinct trade-offs in latency, throughput, area, and power.
The experimental results reveal that while specialized HLS implementations generally offer superior area and power efficiency for lighter workloads such as vector magnitude squaring, the proposed spatial array achieves high efficiency for compute-intensive kernels. Notably, the spatial array demonstrates superior performance across all metrics for matrix-matrix multiplication, even when compared to HLS implementations utilizing an equivalent number of multipliers. Furthermore, the analysis highlights the distinct advantage of the spatial array in a reconfigurable context; a single hardware unit inherently supports all kernels, whereas the specialized approach requires the aggregation of multiple individual cores to achieve equivalent functional coverage. Ultimately, this study confirms that general-purpose systolic architectures can approach the efficiency of specialized cores under specific conditions, offering a promising architectural pathway for the design of agile and scalable next-generation wireless base stations.

\bibliographystyle{IEEEtran}
\bibliography{refs/bibliography}{}

\end{document}